\documentstyle[twoside,12pt]{article}
\newcommand{\ra}{\rightarrow}

\newcommand{\vp}{\varphi}
\newcommand{\vt}{\vartheta}

\normalsize
\begin{document}
\bibliographystyle{plain}

\begin{titlepage}
\begin{flushright}
UWThPh-1994-5\\
January 1994
\end{flushright}
\vfill
\begin{center}
{\Large \bf Gravitating solitons and hairy black holes} \\[40pt]
Piotr Bizo\'{n} \\
Institute of Physics, Jagellonian University,\\
 Cracow, Poland\\
e-mail: bizon@ztc386a.if.uj.edu.pl
\vfill
{\bf Abstract} \\
\end{center}
A brief review of recent research on soliton and black hole solutions
of Einstein's equations with nonlinear field sources is presented and
some open questions are pointed out.

\vfill
\end{titlepage}

\section{Introduction}
During the past few years there has been a lot of interest in soliton
(by which I mean non-singular finite energy stationary solution)
and black hole solutions of \mbox{Einstein's} equations with nonlinear field
sources. In this brief survey I shall outline the most interesting, in
my view, results of these studies\footnote{I tried to make this survey
a complementary
up-date of an excellent review on this subject written two years ago by
Gibbons [1].}. Since my intention is to emphasize the
model-independent aspects of the problem I shall restrict the details of
specific models to a minimum (with one important exception of the
Einstein-Yang-Mills theory which deserves separate analysis).

It is rather surprising that although solitons in various nonlinear
field theories in flat space were intensively investigated in the past,
 the analogous problem in general relativity received little attention until
recently. Presumably this lack of interest in Einstein's equations with
 nonlinear field sources was due to two widely accepted beliefs (apart
from the psychological factor due to the formidable
structure of equations).  First, it has been
thought that the gravitational interaction, being very weak, cannot change
qualitatively the spectrum of soliton solutions of a theory in
which gravity was neglected. Second, according to the so called "no-hair"
conjecture, a stationary black hole is uniquely determined by global charges
given by surface integrals at spatial infinity, such as mass, angular momentum,
and electric (magnetic) charge. It was believed that this property of
black holes, following from the uniqueness [2] and the non-existence [3]
results proven rigorously for various linear field sources,
persists for general matter sources.

The situation has changed radically when it was discovered that
the Einstein-Yang-Mills (EYM) equations admit static non-abelian soliton [4]
and black hole [5] solutions. This came as a surprise because i) neither
the vacuum
 Einstein equations [6], nor pure Yang-Mills equations [7] have soliton
solutions, and ii) "colored" black holes have non-abelian hair which is not
associated with any conserved charge. These unexpected results have put into
question the two beliefs mentioned above and have launched intensive
investigations of Einstein's equations with nonlinear field sources.

 There are several, more or less independent, motivations to study this
 subject. Let us mention some of them. First and most important, this research
 may have serious conceptual implications.
Our understanding of general relativity is to large extent based
on the knowledge of few exact solutions ({\it e.g.} Kerr-Newman,
Friedmann). However, as the results of [4] and [5] showed, the intuitions
based on solutions with linear field sources fail in more
general situations. In this sense the analysis of Einstein's equations with
nonlinear field sources may shed new light on generic
properties of solutions of Einstein's equations.

 Particularly interesting are {\em non-perturbative} effects of gravity, as,
for instance, the existence of gravitational equilibria of non-abelian gauge
fields [4] with finite energy. They result from the cancellation of gauge and
gravitational singularities which
implements the old idea due to Einstein that gravity may regularize ultraviolet
divergencies in field theory. Such non-perturbative phenomena
may have deep consequences at the quantum level.

Second, although we do not have yet a unified quantum theory of all
interactions, we obviously can couple gravity to the standard
model at the classical level. Then, it is natural to expect
that the effects of gravity become important
when the relevant energy scale is close to the Planck scale, {\it i.e.}
in the very early Universe. This fact is notoriously ignored in cosmological
literature. For example, in discussions how inflation
solves the monopole problem the gravitational properties of
monopoles are neglected, which might be a serious omission.

Third, the long time behaviour of perturbed solitons
is closely related to the main unsolved problem in general relativity:
 cosmic censorship hypothesis.

Finally, there is a purely mathematical motivation to study these problems.
In the spherically symmetric case (which is physically most interesting)
 Einstein's equations with nonlinear field sources reduce to dynamical
systems having very rich structure. It seems that modern methods
of bifurcation and critical point theories may be
succesfully applied to these systems.

The rest of this brief review is organized into two sections dealing
with solitons and black holes, respectively. In Section~2
the general properties of gravitating solitons are described.
For weak coupling the perturbative effects of gravity are examined, while
for strong coupling the critical dependence on the coupling constant is
analyzed. Next, the non-perturbative phenomena are discussed, in particular
the Bartnik-Mckinnon solutions of the EYM equations. Section~3 is devoted to
black holes with nonlinear hair and the current status
 of "no-hair" conjecture. This review is concerned with fundamental aspects
of gravitating solitons and hairy black holes and I shall not discuss here
an interesting problem of
possible astrophysical and cosmological relevance of these solutions.

I shall use units in which $c=1$; all other dimensional parameters are
explicitely written.

\section{Globally regular solutions}
Let us consider
 a nonlinear field theory in flat space and suppose that the field
equations have a soliton solution. For the purpose of this review I define
the term soliton to mean a stationary solution which is everywhere
non-singular and has
finite energy\footnote{ I do not require stability, so this meaning is
broader than that usually used in literature.}.
 The following question
arises: What happens when the gravitational interaction is included into the
model? In particular, how does gravity affect the spectrum of soliton
solutions? To study this problem we have to consider Einstein's equations
with a solitonic field as the source.

Before doing this, it is worth recalling
that a theory which satisfies the dominant energy condition and is scale
invariant, cannot have soliton solutions. The reason is that under the scaling
transformation $x \ra \lambda x$, the energy scales as
$E \rightarrow \lambda E$, so
a stationary solution, being the extremum of energy, must have $E=0$. Since the
energy is positive this implies that the solution is trivial. Thus, our flat
space theory, having a soliton, must have a scale
of length, call it $L_0$, which determines the characteristic
size of the soliton (of course there is also a corresponding scale of energy).
The gravitational coupling introduces into the model the additional
dimensional parameter, Newton's constant $G$, which allows to define the second
scale of length $L_g$ (and energy).
Thus typically the Einstein-solitonic-matter
system has two scales of length and the ratio $\alpha=L_g/L_0$
is a dimensionless parameter characterizing the model.
Although my considerations are supposed to be
model-independent, it is helpful to keep in mind a specific model for
illustration. For this purpose I shall use the Einstein-Skyrme model [8,9].
In this model the two scales of length are given by
$L_0=1/ev$ and $L_g=G^{1/2}/e$, where $v$ and $e$ are dimensional
parameters, and $\alpha=G^{1/2} v$ (cf.[8]).\\

{\parindent=0pt\large\bf Weak coupling}\\

For small $\alpha$ one may apply the standard
perturbative argument to prove the existence of gravitating solitons.
The field equations may be expressed schematically in the form
$$
F(f,\alpha) = 0, \eqno(1)
$$
where $F$ is the differential operator
and $f$ denotes the collection of all unknown metric
and matter field variables. Suppose that the
parameters which determine $L_0$ are fixed, hence
 the limit $\alpha \rightarrow 0$
corresponds to switching-off gravity.  By assumption,
 for $\alpha=0$ (without gravity) there exists a soliton solution $f_0$
 satisfying
$$
F(f_0,0) = 0. \eqno(2)
$$
A natural idea to show the existence of
solutions of Eq.(1) is to use the implicit function theorem .
In the present context the appropriate version of the implicit function theorem
is the following. Let $X,Y$ be Banach spaces, $F:X \times R^1 \ra Y$ be a
smooth mapping and let a point $p=(f_0,0) \in X \times R^1$ be a solution
of Eq.(2). Then, if the derivative mapping $D_f F(p)$
is bijective, the theorem guarantees that in the
neighbourhood of $p$ there exists a solution $f(\alpha)$
of Eq.(1), such that $f(0)=f_0$. The derivative  $D_f F$
is just the linearized operator  $\frac{\delta F}{\delta f}$
 one obtains when linearizing $F$.
In the most interesting physical situation the soliton $f_0$ is linearly
stable which means that the eigenvalue problem
$$
\frac{\delta F}{\delta f}(p) \xi = \omega^2 \xi \eqno(3)
$$
has only positive eigenvalues, so  the linear operator
$\frac{\delta F}{\delta f} (p)$
is invertible, and, after choosing appropriate Banach spaces, bijective.
If the spectrum of Eq.(3) is not positive (as it happens for unstable
 solitons), we have to make sure in addition that there are no zero
modes\footnote{In gauge invariant theories
there are pure gauge zero
modes which one has to remove before applying the implicit function
argument. This can be achieved either by fixing the gauge or by working with
the space of gauge orbits.}.
 We conclude therefore that, in general, the soliton solution
persists for sufficiently small $\alpha$.  Since the
gravitational binding energy is negative, the energy of a gravitating
soliton decreases with $\alpha$.
In "everyday" situations $\alpha$ is extremely small, so
the gravitational effects are negligible.

It might seem from the above argument that in the region of weak coupling
nothing spectacular can happen. However, sometimes even for weak coupling
interesting things occur. To see this, notice that I assumed above that
 $\alpha=0$ corresponds to $L_g=0 \;(G=0)$.
However, since $\alpha=L_g/L_0$, the limit $\alpha \ra 0$ may be
obtained in another way, namely by keeping $L_g$
fixed and taking $L_0 \ra \infty$. This results in Einstein's equations
with some truncated matter sources. For example in the Einstein-Skyrme
theory in this limit ($v \ra 0$ with $G$ fixed) the sigma-model term
disappears from the lagrangian. An interesting situation arises when such a
limiting theory, albeit usually unphysical, has a soliton solution. Then, in
 analogy to Eq.(1) we have
$$
\bar{F}(\bar{f},\alpha) = 0, \eqno(1a)
$$
and for $\alpha=0$ there is a solution $\bar{f}_0$ satisfying
$$
\bar{F}(\bar{f_0},0) = 0. \eqno(2a)
$$
The bar over $F$ and $f$ means that although Eqs.(1) and (1a)
are equivalent when $\alpha>0$, the operators $F$ and $\bar F$  are different;
Eq.(1a) is obtained from Eq.(1) by a scaling transformation (depending on
$\alpha$). The Eqs.(2) and (2a) describe different limiting theories;
the procedures of rescaling and taking the limit $\alpha \ra 0$ do not commute.
 If the operator
 $\frac{\delta \bar{F}}{\delta \bar{f}} (p)$ is bijective, the implicit
function theorem guarantees the existence of a solution $\bar{f}(\alpha)$ for
sufficiently small $\alpha$. Then for small $\alpha$ there are two
distinct solutions, $f$ and $\bar f$, which are perturbations of $f_0$ and
 $\bar f_0$, respectively. It should be emphasized that
although the solution $\bar{f}$ is obtained by the perturbative argument,
 its occurence, from the point of view of the original
flat space model, is non-perturbative.  This interesting phenomenon was
observed and described in more detail
 in the Einstein-Skyrme model [8]. It seems that this effect is characteristic
for a certain class of models possessing two scales of length.\\

{\parindent=0pt\large\bf Strong coupling}\\

I have argued above that, as long as the dimensionless coupling constant
$\alpha$ is sufficiently small, a flat-space soliton survives the gravitational
coupling. Will such a gravitating solution persist for large values of
 $\alpha$? The gravitational interaction becomes important when
$\alpha \sim  1$ since then the size of the soliton is of the order of its
Schwarzschild radius. For example the typical size of the skyrmion is
$R \sim 1/ev$ while its mass is $M \sim v/e$, so $GM/R \sim \alpha^2$.
Hence, one might expect that for $\alpha \sim 1$
the gravitating soliton becomes unstable
with respect to the gravitational collapse and
there is a critical value $\alpha_0 \sim 1$ beyond
 which no solitons exist. In fact, this expectation was confirmed  numerically
in several models [8-12]. For example in the Einstein-Skyrme model
$\alpha_0 \approx 0.2$. It is plausible that this behaviour is generic for
stable strongly gravitating solitons\footnote{Unstable gravitating solitons
 may exist for all values of $\alpha$ [13].}.

To understand better the existence of a critical value $\alpha_0$ let us
recall the standard technique of proving existence of solitons in flat
space. Typically one starts from a topological argument showing that the
configuration space has a nontrivial topological structure and splits
into homotopy classes labeled by a winding number (topological charge).
 Then one shows that within a given homotopy class the
energy functional is bounded from below by a positive constant proportional
to the topological charge and therefore there exists a positive lower bound
for energy
$$
E_0 = inf\; E[f]. \eqno(4)
$$
The idea is to prove that this bound is attained by a regular function $f_0$
which is thereby a stationary solution.
Technically this is a three step procedure:
\begin{enumerate}
\item Construct a minimizing sequence $\{f_n\}$, where $f_n$ belong to
 the space $M$ of regular finite energy functions, such that
$\lim_{n \rightarrow \infty} E[f_n]=E_0$. The existence
 of such a sequence is guaranted by Eq.(4).
\item Prove that the sequence $\{f_n\}$ has a limit $f_0 \in M$.
\item Finally, show that $E[f_0] = \lim_{n \rightarrow \infty} E[f_n]$.
This step is nontrivial because usually $E[f]$ is not continuous in $M$.
\end{enumerate}
The proofs of existence of topological solitons along these lines were
given for several models involving non-abelian gauge fields [14].

 Now, let us return to gravitating solitons
and follow the above procedure. First notice that as long as we deal with
asymptotically flat solutions with $R^3$ topology on $t=const$ slices,
the topological arguments are still valid for gravitating solitons (as we
 shall see below in the black hole context the situation may be different).
Similarly, the mass is bounded from below by a positive constant.
However, when one tries to repeat the steps 1-3 in the case of nonzero
 $\alpha$, it turns out that for large $\alpha$
a minimizing sequence $\{f_n\}$ has no limit in $M$. In other words, the
minimizing solution $f_0$ is singular [11]. This shows that one should
be careful
with topological arguments. The topological argument is only a hint for
existence of a solution and has to supplemented by the real proof, which, as
the above example teaches us, might not be a "mere technicality".

The behaviour of gravitating solitons in the strong coupling region is
not yet well understood. In particular, a general argument for the existence
of a critical value $\alpha_0$ is lacking. One might try to approach the
problem along the lines of [15], where some universal upper bounds for the
binding energy of static configurations were derived. In specific spherically
symmetric models another approach seems to be easier. Impose regular initial
 data at $r=0$
 and  show that, when $\alpha$ is large, a local solutions cannot be extended
to a global solution. Actually, this is exactly how such systems are studied
numerically.

A closely related issue is the question of stability of gravitating solitons.
Let us recall that the the static solution $f$ is
said to be linearly stable if all linear perturbations $\delta f(t)$ around it
remain bounded (in a suitable norm) in time. The usual procedure of
investigating linear stability is to assume that $\delta f= e^{i \omega t}
\xi$
, where $\xi$ is time-independent and linearize time-dependent version of
Eq.(1) to get
$$
\frac{\delta F}{\delta f} \xi = \omega^2 W \xi \;,\eqno(5)
$$
where $W$ is some positive weight function. The solution $f$ is unstable if
there exists at least one  eigenmode $\xi$ (satisfying appropriate boundary
conditions) with negative eigenvalue $\omega^2$,
since then $\delta f$ will grow exponentially in time. Otherwise the solution
is linearly stable.
If the flat space soliton is stable then its gravitating counterpart will
also be stable for small $\alpha$, because the eigenvalues $\omega^2$ change
continuously with $\alpha$. The mixed numerical and analytical analysis
 shows that the eigenvalues $\omega^2$ decrease
with $\alpha$ and for $\alpha \ra \alpha_0$ the lowest eigenvalue $\omega^2$
tends to zero. This is not a coincidence, but follows from the general
relationship between zero modes of the linearized operator and bifurcation
points\footnote{ In the presence of symmetries there might be zero eigenvalues
 which have nothing to do with bifurcations (Goldstone modes).}.
This connection is one of the basic results of the catastrophe theory which
allows to study stability problems in the essentially non-dynamical way [16].
For gravitating solitons this fact was first observed in the Einstein-Skyrme
model [8], where at $\alpha_0$ the branch of fundamental gravitating skyrmions
 $f(\alpha)$ merges with another branch of unstable solutions $\bar f(\alpha)$
(discussed in the text below Eq.(1b)).\\

{\parindent=0pt\large\bf Gravitational desingularization}\\

Above I have discussed Einstein's equations with solitonic matter sources.
Now, consider a different situation and suppose that a field source has no
soliton solution in flat space. Can the
 gravitational coupling help in this respect, {\it i.e.} might there exist
globally
regular solutions in the coupled Einstein-non-solitonic-matter theory?
 First, notice that if
the field theory is scale invariant then, according to the remarks above,
the necessary condition for existence of a soliton in Einstein-matter system
is that gravity breaks scale invariance. For this reason the Einstein-massless
scalar or the Einstein-Maxwell equations cannot have a soliton,
 while the Einstein-Yang-Mills (EYM) equations
might have, because the EYM theory has a scale of length given by $\sqrt{G}/e$,
where $e$ is the gauge coupling constant.

Even when a field theory has no
globally regular solutions, it usually will have solutions which
behave correctly at large distances (but are singular at the origin). A typical
example is the monopole solution (electric or magnetic) in electrodynamics.
It should be stressed that the leading asymptotic behaviour at spatial
infinity of such solutions will not be altered by
the gravitational coupling because  gravity decouples from sources at
 infinity. However,  gravity will
modify the short distance behaviour of solutions, and in particular
may regulate their singularities.
This idea is traced back to Einstein who believed that ultraviolet
divergencies in field theory will be eliminated in some unified nonlinear
theory. Let us illustrate such a desingularization phenomenon with
two examples.

Consider a spherical shell of radius $r$ with uniform charge and mass density.
Let $e$ be its total charge and $m_0$ its bare mass. Ignoring gravity, the
energy is given by
$$
m(r) = m_0 +\frac{e^2}{r} \; \eqno(6)
$$
which, of course, diverges as $r$ tends to zero.
Taking into account the gravitational binding energy, this formula is replaced
by
$$
m(r) = m_0 + \frac{e^2}{r} - \frac{G m^2}{r} \;.  \eqno(7)
$$
Solving this equation for $m(r)$ and taking the limit $r \ra 0$, yields
$$
m(r=0) = \frac{e}{\sqrt{G}} \;. \eqno(8)
$$
This heuristic argument can be made rigorous by solving the constraint
equations in the Einstein-Maxwell theory [17].
The result (8) may be viewed as non-perturbative
cancellation of two infinite self-energies: positive electrostatic
and negative gravitational.

To show another example of similar type let us consider a
very simple theory in which gravity is modelled by a massless scalar field
called dilaton (scalar gravity). Dilaton couples in a universal way
to matter (with lagrangian $L_m$) through the term
$e^{-2a\phi}L_m$, where $a$ is the dilaton coupling constant.
 If we take the electromagnetic field as matter, the static Maxwell-dilaton
equations
$$
d( e^{-2a \phi} \star F) = 0, \qquad \nabla^2 \phi + \frac{a}{2}e^{-2a \phi}
 F^2 = 0
\eqno(9)
$$
 have a simple spherical solution with magnetic charge $1/e$ [18]
$$
eF=  d\vt \wedge \sin \vt d\vp, \qquad
a \phi = \ln( 1+ \frac{a}{er} ).  \eqno(10)
$$
Without a dilaton the energy density of the monopole diverges at $r=0$
as $T_{00} \sim 1/r^4$, whereas in the present case $T_{00} \sim e^{-2a\phi}
F^2 \sim 1/r^2$, hence the total energy is  finite (equal to $1/ea$).
Although the total energy of the dilatonic monopole is
finite, the solution (10) is still singular at $r=0$.
Below I shall discuss the EYM theory where gravitational coupling leads to
non-perturbative globally regular solutions.\\

{\parindent=0pt\large\bf Bartnik-Mckinnon solutions}\\

Probably a single most important result in the study of gravitating
solitons was
the discovery of solitons in the EYM system by Bartnik and Mckinnon (BM)
in 1988 [4]. Since then many properties of BM solutions were understood but
 still some questions remain to be answered. Below I discuss briefly the main
features of BM solitons and point out open problems.

The Einstein-YM coupled system is described by the action
$$
S = \int d^4 x \sqrt{-g} \left[\frac{1}{G} R
- {\cal F}^2 \right] \;, \eqno(11)
$$
where $R$ is a scalar curvature and  ${\cal F} = d{\cal A} + e {\cal A}
\wedge {\cal A}$ is the
Yang-Mills curvature of a YM connection ${\cal A}$ which takes values
in the Lie algebra of the gauge group ${\cal G}$. BM considered ${\cal G} =
SU(2)$. There are two dimensional parameters in the theory: Newton's constant
$G$ (of dimension $length/mass$) and the gauge coupling constant $e$
 (of dimension $mass^{-1/2}length^{-1/2}$).

Let us consider static spherically symmetric configurations. It is
convenient to parametrize the metric in the following way
$$
ds^2 = - A^2 N dt^2 + N^{-1}dr^2 + r^2(d\vt^2 + \sin^2 \vt d\vp^2)\;, \eqno(12)
$$
where $A$ and $N$ are functions of $r$.

Assuming that the electric part of the YM field vanishes (actually
this is not a restriction because one can show,
[19], that there are no globally regular static solutions with nonzero
electric field) the
purely magnetic static spherically symmetric $SU(2)$- YM connection can be
written, in the Abelian gauge, as [20]
$$
e{\cal A} = w \tau_1 d\vt + (\cot \vt \tau_3 + w \tau_2) \sin \vt d\vp\;,
\eqno(13)
$$
where $\tau_i$ ($i = 1,2,3)$ are Pauli matrices and $w$ is a function
of $r$.
The corresponding YM curvature is given by
$$
e{\cal F} =  w' \tau_1 dr \wedge d\vt + w' \tau_2 dr \wedge \sin \vt d\vp
 - (1-w^2) \tau_3 d\vt \wedge \sin \vt d\vp\;,
\eqno(14)
$$
where prime denotes derivative with respect to $r$.

Inserting these ans\"{a}tze into the action (11) yields the reduced
lagrangian\footnote{ Strictly speaking, to obtain
(15) one has to add a surface term to the action (11).}
 (where $S=16\pi\int L dt$)
$$
L =  -\int_{0}^\infty \; \left[ \frac{1}{2G}r A'(1-N) + A U \right] dr\;\;,
\eqno(15)
$$
where
$$
U = \frac{1}{e^2}
\left[ N w'{}^2 + \frac{(1-w^2)^2}{2r^2} \right]\;.
\eqno(16)
$$
Variation of $L$ with respect of $w,A$, and $N$ yields the field
equations\footnote{The principle of symmetric criticality
[21] guarantees that the variation
of $S$ within the spherically symmetric ansatz gives the correct
equations of motion.}
$$
(A N w')' +  \frac{1}{r^2} A w(1 - w^2) = 0\;, \eqno(17)
$$
$$
\left[ r (1-N) \right]' = 2 G U\;,
\eqno(18)
$$
$$
A' =  \frac{2G}{e^2 r} A w'{}^2 \;.
\eqno(19)
$$
Note that, using Eq.(19), $A$ may  be eliminated from Eq.(17).

These equations have three explicit abelian solutions. The first two
are the vacua $w=\pm 1$, $A=N=1$ with zero mass (or Schwarzschild
$A=1, N=1-2GM/r$). The third solution
$$
w=0\;, \qquad A = 1\;, \qquad
N = 1 - \frac{2GM}{r} + \frac{G}{e^2 r^2}  \eqno(20)
$$
 describes the Reissner-Nordstr\"{o}m black hole with mass $M$ and
magnetic charge $1/e$.

In order to construct non-abelian solutions
which are globally regular one  has to
impose the boundary conditions which ensure regularity at $r=0$ and
asymptotic flatness. The asymptotic solutions of Eqs.(17,18)
satisfying these requirements are
$$
\pm w = 1 - b r^2 + O(r^4) \;, \qquad N = 1 -4 b^2 r^2 + O(r^4) \eqno(21)
$$
near $r=0$, and
$$
\pm w = 1 - \frac{c}{r} + O(\frac{1}{r^2}) \;,
\qquad N = 1 - \frac{d}{r} + O(\frac{1}{r^2}) \eqno(22)
$$
near $r=\infty$. Here $b,c,d$ are arbitrary constants.

 Notice that, for the asymptotic behaviour (22), the
radial magnetic curvature, ${\cal B}_r = \tau_3 (1-w^2)/r^2$, falls-off
as $1/r^3$, and therefore all globally regular solutions have
zero YM magnetic charge.

Bartnik and Mckinnon gave strong numerical evidence that there exists a
countable sequence of
 initial values $b_n$ ($ n \in N$) determining globally
regular solutions $w_n,A_n,N_n$. The index $n$ labels the number
of nodes of the function $w_n$. The sequence of masses $M_n$ of these
solutions increases with $n$ and tends to $M_{\infty}=1$ for $n \ra \infty$
(the unit of mass is given by $1/G^{1/2}e$). If one defines the
"fine structure" constant $\gamma^2 = \hbar e^2$, then the mass
of BM solitons is of the order
$ m_{Pl}/\gamma$ while their effective size is of the order $l_{Pl}/
\gamma$. Notice that the classical treatment of such objects
 makes sense physically only
if $\gamma$ is small since then
the Compton radius is much less than the effective radius of
BM solitons.

The BM discovery raised a number of questions. Let us enumerate the
most important ones:
\begin{itemize}
\item {{\bf Existence}:}
The first rigorous proof of existence of BM solitons was given
by Smoller {\it et al}. [22].  Another proof was recently presented by
 Breitenlohner {\it et al}. [23]. Both proofs use the methods
of dynamical system theory to analyze the a priori behaviour of orbits
starting with regular initial data at $r=0$ and find that there exists
a countable family of connecting orbits (separatrices) which correspond
to the BM solutions. Both proofs leave open the question of
 uniqueness of BM solutions (amongst globally regular solutions of
Eqs.(17-19)).
\item{{\bf Stability}:}
 The linear stability analysis of BM solutions
was carried out by \mbox{Straumann} and Zhou [24].
 They derived an
eigenmode equation, such as Eq.(5), which governs the time evolution of
small perturbations around the BM solitons and demonstrated that
it has at least one negative eigenvalue (unstable mode).
Later [25] they also analyzed the long time behaviour of perturbed
 BM solitons and argued that, depending on an initial perturbation, they
either disperse to infinity or collapse to form the Schwarzchild black
hole.
\item{{\bf Raison d'\^{e}tre}:} In the study of nonlinear equations
perhaps even more important than proving existence of a solution is
to understand the reason for its existence.
  The existence proofs mentioned above do not really
 explain what are the essential features of the $SU(2)$-YM
gauge group which are responsible for the existence of BM solutions and
their basic properties, such as discreteness and instability.
Soon after the discovery of BM solutions P. Mazur suggested
(private communication, 1990)
that they should be regarded as gravitational sphalerons.
Let us recall that sphaleron is a saddle point solution whose existence
follows from the Lusternik-Snirelman mini-max (mountain pass) construction
 applied to the
space of non-contractible loops (or paths joining
topologically non-equivalent vacua) of field configurations. Sphalerons were
first discovered
by Taubes and Manton in spontanously broken gauge theories (YM-Higgs) [26],
but they exist in many
other models which possess non-contractible loops in configuration space [27].
 The first published attempt [28] to interpret the
BM solutions as sphalerons was unsuccesful, because the authors considered
loops with too weak decay condition for the YM connection
which led to a disaster of having the
infimum of energy over all loops equal to zero ({\em cf.} footnote 9).

Recently Sudarsky and Wald (SW) [29] proposed,
in the language of Hamiltonian formulation of general relativity,
a very interesting modification of the original Taubes-Manton argument.
In the case of
$n=1$ BM solution, the SW argument is, in essence, equivalent to the mini-max
procedure for paths joining two topologically inequivalent vacua.
However, in contrast to the mini-max construction, the SW mechanism can be
naturally extended to account also for the existence of $n>1$ BM solutions.
The SW argument got support by the discovery of solutions similar
to BM solutions in another theory involving $SU(2)$-YM field [18].
It is very unlikely that the SW argument it its present form can be converted
into a genuine proof. However, in the spherically symmetric case the situation
simplifies considerably and seems to be tractable rigorously. I shall come back
to this point when discussing the YM field on Schwarzschild background.

\item {{\bf Spectrum of masses}:}  The discrete spectrum of masses of BM
solutions
has an accumulation point (upper bound) at $M=1/(eG^{1/2})$. This follows from
the fact that for $n \ra \infty$ the BM solutions tend (non-uniformly)
 to the extremal Reissner-Nordstr\"{o}m solution with magnetic
charge equal to $1/e$.
Can one construct an inequality $M \leq 1/(eG^{1/2})$ which is saturated by the
limiting solution?
\item {{\bf Spherical symmetry}:} In the EYM system the Birkhoff theorem is
not valid since a spherically symmetric configuration need not be static.
 Is the converse result true, {\it i.e.} does staticity imply spherical
symmetry? Showing this
 would be the first step towards the proof of
uniqueness of BM solutions (of static $SU(2)$-YM equations).
 One could try to attack this problem using the technique of
Bunting and Masood-ul-Alam [30] of finding a conformal transformation of
 spatial
3-metric into a zero-mass metric with non-negative scalar curvature.
As far as I am aware (W. Simon, private communication) this approach encounters
serious technical difficulties.
A closely related question: are there stationary axisymmetric analogues of BM
solutions?
\item {{\bf Desingularization}:} The BM solutions are non-perturbative in the
gravitational constant $G$ -- nota bene in the EYM theory
there is no dimensionless parameter which might serve as a perturbative
parameter (changing the coupling constants $e$ and $G$ only changes the
 scale).
Although the flat-space static YM equations have no globally regular
 solutions [7], they possess
 spherically symmetric solutions which at large distances behave
similarly to BM solutions. These are solutions of Eq.(17) with $A\equiv
N\equiv 1$ [31]. Their asymptotic behaviour at infinity is given by Eq.(22),
 whereas near $r=0$ they behave as
$$
w \sim \sqrt{r} \sin(\frac{\sqrt{3}}{2} \ln{r}+const)\;.  \eqno(23)
$$
so the YM curvature $\cal F$ is singular at $r=0$ and the total energy
is infinite.
I believe that the BM solutions may be viewed as the regularized (by gravity)
version of these singular flat-space solutions. In a sense
the proof of \mbox{Breitenlohner} {\it et al.} implements this idea, because
they start from the phase portrait of Eq.(17) in flat space and analyze how
 it is modified by gravity. It would be of interest to pursue this idea
further. One possible direction is to search for exact form of BM solutions,
since having it we probably could see how the scale of length $G^{1/2}/e$
provides a cut-off for singular behaviour (23). Recent results in the
 EYM-dilaton theory (Bizo\'{n} and Chmaj, work in progress) suggest that the
hope of finding an analytical expression for BM solutions is not unrealistic.
\end{itemize}

\section{Black holes}
Let us now turn to another interesting class of asymptotically flat
solutions, namely black holes. We restrict attention to the
region outside the horizon. Then, in the stationary case,
 Einstein's equations reduce
 to the elliptic boundary value problem
between horizon and infinity.

There are more black hole solutions than globally regular ones.
This is due to  two important differences between these two cases.
First, stationary globally regular solutions are extrema of mass, while
 stationary black holes extremize the mass provided that the area of
the horizon is fixed (and the angular momentum is fixed, if there is
nonzero rotation)\footnote{In both cases there might be additional
contributions to the variation of mass coming from other extensive
parameters (such as {\it e.g.} charge). Mass is extremized at stationary
solutions if these parameters are fixed [29].} [29].
This follows from the first law of black hole dynamics
$$
G\delta M = \frac{1}{8 \pi}\kappa \delta A + \Omega \delta J \;, \eqno(24)
$$
where $M,\kappa,A,J$, and $\Omega$ are the black hole mass, surface gravity,
area, angular momentum, and angular velocity of the horizon.
In other words
the presence of horizon introduces the boundary condition which breaks
scale invariance. This allows for nontrivial black hole solutions in scale
invariant theories, as for instance the Schwarzschild solution.

Second, solutions which are singular in flat space
may give rise to
 perfectly regular hair on black holes if the singularities are hidden
behind the horizon. The Reissner-Nordstr\"{o}m black hole is a typical
example.

In analogy to the discussion of globally regular solutions we can distinguish
several situations depending on scaling properties. The first possibility is
that the model is scale invariant, and the scale is given solely by the area of
the horizon. In this case we have a continuous family of black holes, such as
Schwarzschild or Reissner-Nordstr\"{o}m solutions.

The second possibility is that the corresponding flat space model has
a soliton and a scale $L_0$. The coupling of
gravity gives a second scale $L_g$ and the presence of horizon provides a
third scale $r_H$,
where $r_H$ is the radius of the horizon (for simplicity I consider
spherically symmetric case). Thus, in analogy to Eq.(1),
 we can write the field equations
in the form
$$
F(f,\alpha,\beta) = 0 \;, \eqno(25)
$$
where $\alpha=L_g/L_0$ and $\beta=r_H/L_0$. Note that the status of
parameters $\alpha$ and $\beta$ is different -- $\alpha$ is a combination
of coupling constants while $\beta$ is determined by the boundary condition
(in the nonspherical case it is even not clear how to define $\beta$).
Nevertheless, as long as $r_H>0$, by rewriting the equations in the
variable $x=r/r_H$, $\beta$ is eliminated from the boundary condition and
appears as a second (apart from $\alpha$) dimensionless parameter in the
equations .
It is very convenient to think about Eq.(25) as describing black holes when
$\beta>0$, and globally regular solutions when $\beta=0$. That is, we
require $f$ to belong to a space $X$ of asymptotically flat configurations
which for $\beta>0$ satisfy the black-hole boundary conditions at $r_h$, while
 for $\beta=0$ (when black hole becomes singular since the horizon shrinks to
zero)  are regular at the origin. The advantage of working with $X$ is that
the flat-space solution $f_0$, satisfying $F(f_0,0,0)=0$, belongs to $X$ and
may be used as the basis of the perturbative argument. Such an argument was
proposed by Kastor and Traschen [32] who argued, using the implicit function
theorem, that for sufficiently small $\alpha$ and $\beta$ in the neighbourhood
of $f_0$ there exist black hole solutions of Eq.(25). Although their heuristic
argument is probably basically correct, it certainly should be pursued
by more precise analysis which would take care of some technical subtleties,
such as the effects of degenerate ellipticity at $r_H$. Anyhow, this reasoning
is strongly supported by numerical results in several models having
soliton solutions, where black hole counterparts with small horizon
(sometimes referred to as black holes inside solitons) were found
[8-12,33]. When $\beta \ra 0$ these black holes tend to the corresponding
globally regular solutions.

It turns out that such black holes exist only in a bounded
region of the $(\alpha,\beta)$-plane. As was shown before, for $\beta=0$
there is a maximal value $\alpha_0$ beyond which no globally regular
solutions exist. Numerical and analytical analyses show that
when $\beta$ grows, the critical value $\alpha_0(\beta)$ monotonically
decreases and goes to zero for some $\beta_0$ [8-12,33].
Thus there is an upper bound for the radius of horizon. The heuristic
explanation is that when the horizon gets larger, more and more of solitonic
hair is swallowed and finally, when the radius of the horizon is
comparable to the size of the soliton ($\beta \sim 1$), the whole soliton
disappears inside the black hole (no "cloud" outside the horizon is left).
The technical reason of non-existence of black holes with solitonic hair
for large $\beta$ is the following. The standard no-hair proof (see
papers by Bekenstein [3]) consists in constructing an identity which has
the form of a divergence equaling a non-negative definite quantity.
After integrating this identity between horizon and infinity the divergence
term vanishes, hence the integral of non-negative quantity is zero which
implies that the field (hair) is trivial. For a nonlinear source one can also
construct a similar divergence identity but usually its right side has both
positive and negative contributions, hence the proof fails. However in some
cases one can
show that when $\beta$ is large enough the positive terms take over so
the right side of the divergence identity is non-negative and the no-hair
proof goes through.

Note that the limiting case $\alpha=0$ corresponds to the decoupling of
 gravity. Thus to find a maximal value $\beta_0$ it is sufficient to
study solitonic fields on fixed Schwarzschild background. For skyrmionic
hair this was done in [34].

The third possibility is that the flat space model has no solitons,
and gravity brings a scale $L_g$. Then the field equations have the
form $F(f,\gamma)=0$, where $\gamma=r_H/L_g$. This situation arises in the EYM
theory so let us now consider black holes in this model. As above, it is
 convenient to
treat the limit $\gamma=0$ as corresponding to a regular solution.
As I have already described, in this case there is a countable family
of BM solutions; let us denote them by $f_n^0$ (superscript $0$ refers
to $\gamma=0$). For sufficiently small $\gamma$ we can repeat Kastor
 and Traschen's heuristic argument to predict the existence of
a countable family of black holes $f_n(\gamma)$, which are perturbations of
the corresponding BM
solutions. These solutions, so called colored black holes (CBHs), were found
numerically independently by several authors [5] soon after the
 BM discovery. Recently, the existence of CBHs
was established rigourously [22,23]. Both numerical and
analytical results show that there is no upper bound for $\gamma$,
{\it i.e.} for every $\gamma$ there is a countable family of CBHs,
$f_n(\gamma)$.
It should be stressed that CBHs with different $\gamma$ are essentially
different; in particular they are {\em not} related by any simple scaling
(contrary to some statements scattered in literature). When
$\gamma \ra \infty$ the YM and gravitational fields decouple and we end up
with the YM field on fixed Schwarzschild background. This model is extremely
 simple, nevertheless it seems to  contain all the essential features which
are responsible for the existence of BM-type solutions.
For this reason it is an ideal ground for converting heuristic arguments
(such as the SW argument) into rigorous results.
To my knowledge this model was not described in literature so I take this
opportunity to give some details.\\

{\parindent=0pt\large\bf{SU(2)-YM on Schwarzschild background}}\\

Let us rewrite Eqs.(17-19) using a dimensionless coordinate $x=r/r_H$ and
take the limit $\gamma=e r_H/G^{1/2} \ra \infty$. In this limit
the right side of Eqs.(18,19) vanishes, hence they are solved by the
Schwarzschild metric
$$
N = 1 - \frac{1}{x}, \qquad A =1\;,  \eqno(26)
$$
and therefore Eq.(17) reduces to
$$
\left((1-\frac{1}{x})w'\right)' + \frac{1}{x^2} w(1-w^2) = 0\;. \eqno(27)
$$
This equation follows from the variation of the energy functional
$$
E[w] = \int_1^{\infty} \left[ (1-\frac{1}{x}) w'^2 + \frac{1}{2 x^2} (1-w^2)^2
\right] dx\;. \eqno(28)
$$
It was shown first numerically and recently analytically (Bizo\'{n},
unpublished) that Eq.(27) has a countable family of solutions $\{w_n\}$
($n \in Z$) which are regular for $x \geq 1$. The first member of this family
is the vacuum $w_0=1$, for which the energy functional attains the
global minimum $E[w_0]=0$. For a discrete set of initial values $w_n(1)\in
(0,1)$ there exist essentially non-abelian solutions $w_n$ which oscillate
$(n-1)$-times within a strip $(-1,1)$ and tend to $(-1)^n$ for $x \ra \infty$.
The discrete spectrum of energies $E_0=0,E_1\approx 0.4795,E_2\approx 0.4994,
E_3\approx 0.4999$,... has an accumulation point at $E_{\infty}=1/2$.
 This follows
from the fact that for $n \ra \infty$ the solutions $w_n$ tend pointwise (on
any finite interval $[1,x_0)$) to the abelian solution $w_{\infty}=0$ with
energy $E[w_{\infty}]=1/2$. Using Eq.(27) one can easily show that on-shell
the energy is given by
$$
E[w] = \int_1^{\infty} \frac{1}{2x^2} (1-w^4) dx \;, \eqno(29)
$$
which proves that the solution $w_{\infty}=0$ has maximal energy.

For completeness let us note
that the solution $w_1$ is known analytically [35]
$$
w_1 = \frac{c-x}{x+3(c-1)}\;,  \eqno(30)
$$
where $c=(3+\sqrt{3})/2$.

Notice that our system has  $Z_2$ symmetry, hence along with $\{w_n\}$
there is a corresponding
family of reflected solutions $\{-w_n\}$. The limiting solution
$w_{\infty}$ is a fixed point of reflection.

The Hessian of the energy functional is given by
$$
\delta^2 F(w) (\xi,\xi) = \int_1^{\infty} \left[(1-\frac{1}{x})\xi'^2
+ \frac{1}{x^2}(3 w^2 -1 ) \xi^2 \right] dx \;.  \eqno(31)
$$
A very interesting feature (found numerically) is that the Hessian
evaluated at a solution $w_n$ has exactly $n$ negative eigenvalues.
This property is a strong hint for a mechanism which gives rise to the
family $\{w_n\}$. Before discussing it let me
 make the following clarifying remark.

{\em Remark.} The key feature of the $SU(2)$-YM gauge group is the existence
of large gauge transformations, {\it i.e.} topologically inequivalent cross
sections
of the YM bundle. They arise as follows. Choose the gauge $A_0=0$ and consider
the spatial YM potential $A_i$ on the $t=const$ slice $\Sigma$ of $R^3$
topology. The pure gauge configurations have the form $A_i = \partial_i U
 U^{-1}$, where $U$ is a map from $\Sigma$ to the group manifold $SU(2)$.
Demanding that $U \ra 1$ (sufficiently fast\footnote{It is essential that
$r \partial_i U \ra 0$ (and therefore $r A_i \ra 0$) as $r \ra \infty$.
Otherwise the YM bundle is topologically trivial, as one can easily demonstrate
by constructing curves of connections (with $A_i=O(1/r)$ decay) joining
large gauge transformated vacua. I thank R. Bartnik for pointing this out to
me.}),
 this becomes a homotopically
nontrivial map $U: S^3 \ra S^3$, classified by integer winding numbers.
Thus, there is a countable set of topologically inequivalent vacua which
cannot be continuously deformed into one another. This fact is the basis for
the Taubes-Manton mini-max construction of sphalerons and for the
 Sudarsky-Wald construction of BM solutions. However, in the
 black hole setting the situation is different. The point is that now the
 space $\Sigma$ has topology
 $S^2 \times R^1$ and the mapping $U: \Sigma \ra S^3$ is topologically trivial
(it is not possible to compactify $\Sigma$ by demanding $U$ to be identity on
 the horizon).
\paragraph{}
Although the topology of the $SU(2)$-YM field  on the Schwarzschild
background is trivial, the presence of two distinct vacua and $Z_2$ symmetry
seems to be sufficient for making the mini-max construction. Let $\Gamma$ be
a space of
 functions $w$ for which the energy functional (28) is finite. Consider all
paths in $\Gamma$ connecting $w_0$ and $-w_0$. Since $w_0$ and $-w_0$ are
global minima of energy, on each path there is a point of maximal energy.
The infimum over these maximal energies gives a saddle point (with exactly
one unstable mode), which we identify as $w_1$.
 It is tempting to repeat this
procedure using $w_1$ and its reflection $-w_1$. However, this fails because
there are paths joining $w_1$ and $-w_1$ which go below the energy level
$E[w_1]$. One could try to remedy this obstacle by defining a space $\Gamma_1$
consisting of all maxima on paths joining $w_0$ and $-w_0$, and making mini-max
on $\Gamma_1$. Unfortunately, it is very doubtful that $\Gamma_1$ will be
sufficiently well-behaved to allow such construction\footnote{I thank R. Wald
for pointing this out to me.}, in particular almost
surely $\Gamma_1$ won't be a surface in $\Gamma$.
Recently R. Wald suggested (private communication) that to obtain higher $n$
solutions one should apply mini-max method not for paths but for higher
dimensional objects, such as $n$-spheres which are invariant under reflections.

Another very interesting approach was pursued by O. Popp (unpublished).
The idea is to show that the energy
functional is a Morse function on $\Gamma\setminus w_{\infty}$, {\it i.e.}
 that all critical points
of $E[w]$ (except $w_{\infty}$) have finite index (in the language
of Morse theory, index =
number of negative eigenvalues of the Hessian). To prove this one approximates
$E[w]$ by some truncated functional $F_{\lambda}[w]$, defined by certain
integral over the interval
$[1,\lambda]$. For $\lambda$ close to one, $F_{\lambda}[w]$ has only three
critical points: $\pm w_0=1$ and $w_{\infty}=0$ and all these
 points have zero index.
When $\lambda$ increases new critical points appear. At each bifurcation point
$\lambda_n$ the solution $w_{\infty}$ "sheds" a pair of new solutions
$(w_n,-w_n)$, and at the same time it acquires one additional
negative eigenvalue ({\it i.e.} the index of $w_{\infty}$ increases by one).
Near the bifurcation point $\lambda_n$ the
solution $w_n$ is close to $w_{\infty}$, and therefore it has the same index
(equal to $n$). As $\lambda$ increases the index of $w_n$ remains constant
(this fact is not yet proven rigorously but there is strong evidence that it
is true).
Showing that $F_{\lambda}[w] \ra E[w]$ for $\lambda \ra \infty$ concludes
Popp's argument.\\

{\parindent=0pt\large\bf{No-hair conjecture}}\\

The "no-hair" conjecture belongs to the folklore of general relativity
since the late sixties (for a list of references, see [36]). The conjecture
 concerns
the possible forms of stationary black holes. The idea is that the only
classical degrees of freedom of black holes are those corresponding to
non-radiative multipole moments, because "everything that can be radiated away
will be radiated away" (cf.[37]). For a massless bosonic field with spin $s$,
radiative multipoles have moments $l \geq s$, hence for pure gravity ($s=2$)
the conjecture
allows for a monopole (mass) and a dipole (angular momentum) degrees of
freedom. If there is an electromagnetic field ($s=1$), the electromagnetic
monopoles (electric and magnetic charge) are also allowed. The conjecture
exludes the massless scalar field ($s=0$) and all massive fields, because
for these fields all multipoles are radiative.

At the time when the "no-hair" conjecture was formulated this picture was
perfectly consistent with the fact that the only known stationary black
hole was the Kerr-Newman solution and nobody doubted that this was a unique
stationary black hole solution of the Einstein-Maxwell equations (although this
fact was established rigorously later on [2]). Morever, the conjecture was
supported by several no-go results which showed that
such fields as massless and massive
scalar, massive vector or spinor cannot reside on stationary black holes [3].

The "no-hair" conjecture was formulated rather vaguely and therefore admits
many interpretations. Nowadays most people agree that it should be meant as a
statement concerning the uniqueness of stationary black holes. Let us adopt
this viewpoint and try to formulate the "no-hair" conjecture more
precisely\footnote{I am concerned here with what the "no-hair" conjecture
 really means,
and not with mathematical assumptions underlying the uniqueness theorems
({\it cf.} [38]).}.
 For this purpose let me define a term global charge to mean
a conserved charge associated with a massless gauge field, which is given
by surface integral at spatial infinity. Mass, angular momentum, and electric
and magnetic charges are examples of global charges.
Then the strongest variant of "no-hair" conjecture is
\paragraph{}{\em Version 1.} A stationary black hole is uniquely described
by global charges.
\paragraph{}
For a long time no counterexample to this version was known. However,
recent discoveries of various black
holes with nonlinear matter fields made clear that one cannot speak about
 uniqueness without specifying the sources. To see this consider an
example of Einstein-Maxwell-dilaton theory. In this model
 Gibbons found very interesting black hole solutions [39]. In the
so called string-inspired case (where
the dilaton coupling constant $a$ is equal to $\sqrt{G}$), the
electrically charged solution is given by
$$
ds^2 = -(1-\frac{2GM}{r}) dt^2 + (1-\frac{2GM}{r})^{-1} dr^2
    + r (r-\frac{Q^2}{M}) d\Omega^2
$$
$$
    F = e^{2\sqrt{G}\phi}\,\frac{Q}{r^2}\, dt \wedge dr \qquad \qquad
 e^{2\sqrt{G}\phi} = 1 - \frac{Q^2}{Mr}  , \eqno(32)
$$
where $\phi$ is a dilaton. It was shown recently that this (and the
analogous magnetically charged solution) is a unique
static black hole solution in this model [40]. However, the dilaton field
is not associated with any conserved charge, and therefore at infinity this
solution is indistinguishable from the Reissner-Nordstr\"{o}m solution
with the same mass and charge. Thus an observer at infinity would not be able
to determine a solution by measuring global charges, unless she knows
the matter content of the world. This leads us to the following modification:
\paragraph{}{\em Version 2.} {\em Within a given model} a stationary black
hole is uniquely determined by global charges.
\paragraph{}

Although in many cases, such as the electrovacuum, this version is actually a
 rigorous theorem, there
are theories in which it is false. For example,
as I discussed above, in the EYM theory there is a countable family
of colored black holes. They carry no charge so their non-abelian field
leaves no imprint at infinity.
 For a given
mass (which is the only global parameter) there are infinitely many CBHs with
 different areas of the horizon  which clearly
 violates {\it Version 2}.  However, this is not a physically serious
counterexample, because CBHs are
 unstable [41] - under small perturbation they loose non-abelian hair
and decay into Schwarzschild (for such unstable hair the word wig seems to
be more appropriate). We should therefore add to {\it Version 2} the
following qualification:
\paragraph{}{\it Version 3.} Within a given model a {\em stable} stationary
 black hole is uniquely determined by global charges.
\paragraph{}
Of course in the EYM theory this unique stable solution is Schwarzschild.
Unfortunately, there are models which
violate even this version [8,33]. For example,
in the Einstein-Skyrme model, for a given mass there are three black holes,
 two stable and one unstable. One of the stable black holes is Schwarzschild
while another one has skyrmionic hair. Since the skyrmionic hair is not
connected to any conserved charge (and morever is topologically trivial),
the existence of these two stable black holes provides a counterexample
to  {\em Version 3}.

There seems to be no further possibility of relaxing {\em Version 3} to
accomodate such cases as the Einstein-Skyrme model. We conclude therefore
that there is no universally valid formulation of the "no-hair" conjecture.

\section*{Acknowledgements}
 I am indebted to P. Aichelburg, R. Bartnik,
T. Chmaj, O. Popp, A. Staruszkiewicz, and especially R. Wald
for discussions and remarks. Most of my research in the area which is reviewed
here was done when I was
visiting the Institut f\"{u}r Theoretische Physik and the Erwin
Schr\"{o}dinger Institut f\"{u}r Mathematische Physik in Vienna;
I am grateful to Professors P. Aichelburg and  W. Thirring for hospitality.
 This work was supported in part by the Fundacion Federico.

\end{document}